\documentclass[a4paper,12pt]{article}
\usepackage{indentfirst}
\usepackage{amsfonts,amsmath,amssymb,bm,a4wide,graphicx,cite,makeidx,multicol}

\begin{document}

\title{Spin light of electron in matter}

\author{Alexander Grigoriev~$^{a,b}$ \footnote{ax.grigoriev@mail.ru}, \
        Sergey Shinkevich~$^a$ \footnote{shinkevich@gmail.com}, \
        Alexander Studenikin~$^{a,b}$ \footnote{studenik@srd.sinp.msu.ru},
        \\
        Alexei Ternov~$^c$ \footnote {a\_ternov@mail.ru}, \ Ilya Trofimov~$^a$ %\footnote {a\_ternov@mail.ru}
        \\
   \small {\it $^a$Department of Theoretical Physics,}
   \\
   \small {\it Moscow State University, 119992 Moscow,  Russia }
   \\
   \small {\it $^b$Skobeltsyn Institute of Nuclear Physics,
   Moscow State University, 119992 Moscow, Russia}
   \\
   \small{\it $^c$Department of Theoretical Physics,}
   \\
   \small {\it Moscow Institute for Physics and Technology,
   141700 Dolgoprudny, Russia }}

\date{}

\maketitle

\abstract{ We further generalize the powerful method, which we have recently developed for
description of the background matter influence on neutrinos, for the case of an electron moving in
matter. On the basis of the modified Dirac equation for the electron, accounting for the standard
model interaction with particles of the background, we predict and investigate in some detail a new
mechanism of the electromagnetic radiation that is emitted by moving in matter electron due to its
magnetic moment. We have termed this radiation the \ ``spin light of electron" in matter and
predicted that this radiation can have consequences accessible for experimental observations in
astrophysical and cosmological settings.}

% ----------------------------------------------------------------
\section{Introduction}
In a series of our papers \cite{StuJPA06, StuTerPLB05, GriStuTerPLB05,
GriStuTerPAN06} we have developed a rather powerful method of investigation of
different phenomena that can appear when neutrinos and electrons move in the
background matter.  The method discussed is based on the use of the modified
Dirac equations for the particles wave functions, in which the correspondent
effective potentials, that account for the matter influence on particles, are
included. It is similar to the Furry representation \cite{FurPR51} in quantum
electrodynamics, widely used for description of particles interactions in the
presence of external electromagnetic fields. In \cite{StuTerPLB05,
GriStuTerPLB05, GriStuTerPAN06, StuJPA06} we apply the discussed method for
elaboration of the quantum theory of the \ ``spin light of neutrino" in matter.
The spin light of neutrino in matter ($SL\nu$), one of the four new phenomena
studied in our recent papers \cite{StuNPB05},  is a new type of electromagnetic
radiation that can be emitted by a massive neutrino (due to its magnetic
moment) when the particle moves in the background matter. Within
quasi-classical treatment the existence of this radiation was proposed and
studied in \cite{LobStuPLB03_DvoGriStuIJMP05}, while the quantum theory of this
phenomenon was developed in \cite{StuTerPLB05, GriStuTerPLB05, GriStuTerPAN06,
StuJPA06, LobPLB05}.

It has been shown \cite{StuJPA06} how the approach, developed at first for
description of a neutrino motion in the background matter, can be spread for
the case of an electron propagating in matter. The modified Dirac equation for
an electron in matter has been derived \cite{StuJPA06} and on this basis we
have considered the electromagnetic radiation that can be emitted by the
electron (due to its magnetic moment) in the background matter. We have termed
this radiation as the \ ``spin light of electron" in matter. It should be noted
here that the term \ ``spin light" was introduced in \cite{ITernSPU95} for
designation of the particular spin-dependent contribution to the electron
synchrotron radiation power.

\section{Modified Dirac equation for electron in matter}
Let us consider an electron having the standard model interactions with
particles of electrically neutral matter composed of neutrons, electrons and
protons. Indeed, we account below only for the neutron component that can be
used as an abrupt model for modelling a real situation existed when electrons
move in nuclear matter of a neutron star (see, for instance,
\cite{KusPosPLB02}). We suppose that there is a macroscopic amount of the
background particles in the scale of an electron de Broglie wave length. Then
the addition to the electron effective interaction Lagrangian is
\begin{equation}\label{Lag_f_e}
\Delta L^{(e)}_{eff}=\widetilde{f}^\mu \Big(\bar e \gamma_\mu
{1-4\sin^{2}\theta_{W}+\gamma^5 \over 2} e \Big),
\end{equation}
where the explicit form of $\widetilde f^{\mu}$ depends on the background
particles densities, speeds and polarizations. The modified Dirac equation for
the electron wave function in matter is \cite{StuJPA06}
\begin{equation}\label{new_e}
\Big\{ i\gamma_{\mu}\partial^{\mu}-\frac{1}{2}
\gamma_{\mu}(1-4\sin^{2}\theta_{W}+\gamma_{5}){\widetilde f}^{\mu}-m_e
\Big\}\Psi_{e}(x)=0,
\end{equation}
where for the case of electron moving in the background of neutrons
\begin{equation}\label{f2_e}
\widetilde{f}^{\mu}=\frac{G_{F}}{\sqrt{2}}(n_n,n_n{\bf v}).
\end{equation}
Here $n_n$ is the neutrons number density and $\mathbf v$ is the
speed of the reference frame in which the mean momentum of the
neutrons is zero.

The solution of equation (\ref{new_e}) can be found in analogy with the case of
neutrino (for details see \cite{StuTerPLB05,GriStuTerPLB05}). The complete set
of the electron wave functions in the matter is:
\begin{equation}\label{Spinor_AB_e}
%\begin{array}{c}
\Psi_{\varepsilon, {\bf p},s}({\bf r},t)={\displaystyle
\frac{e^{-i( E_{\varepsilon}t-{\bf p}{\bf r})}}{2L^{3/2}}}
\left(%
\begin{array}{c}{\sqrt{1+\frac{m_e}{ {E}_{\varepsilon}-c\alpha_n m_e}}} \ \sqrt{1+s\frac{p_{3}}{p}}
\\
{s \sqrt{1+ \frac{m_e}{ {E}_{\varepsilon}-c\alpha_n m_e}}} \
\sqrt{1-s\frac{p_{3}}{p}}\ \ e^{i\delta}
\\
{  s\varepsilon\sqrt{1- \frac{m_e}{ {E}_{\varepsilon}-c\alpha_n
m_e}}} \ \sqrt{1+s\frac{p_{3}}{p}}
\\
{\varepsilon\sqrt{1- \frac{m_e}{ {E}_{\varepsilon}-c\alpha_n
m_e}}} \ \ \sqrt{1-s\frac{p_{3}}{p}}\ e^{i\delta}
\end{array}
\right),
%\\
%\\
%%\end{equation}
%%\begin{equation}\label{AB}
%%    A=\frac{m_e}{ E_{\varepsilon}-c\alpha_n m_e},
%    A={\displaystyle \frac{m_e}{ {E}_{\varepsilon}-c\alpha_n m_e}},
%     \quad B=s{\displaystyle \frac{p_{3}}{p}}, \quad
%     \tilde{E}_{\varepsilon}={E}_{\varepsilon}-c\alpha_n m_e,
%\end{array}
\end{equation}
where $L$ is the normalization length,
$\delta=\arctan({p_2}/{p_1})$ and the so called matter density
parameter $\alpha_n$ is defined as
\begin{equation}\label{alpha_n}
 \alpha_n=\frac{{G}_{F}}{2\sqrt{2}}\frac{n_n}{m_e}.
\end{equation}
The electron energy in matter
\begin{equation}\label{E_e}
    E^{(e)}_{\varepsilon}=\varepsilon {\sqrt{{\bf p}^{2}\Big(1-s \alpha_n
\frac{m_e}{p}\Big)^{2}+{m_e}^2} +c \alpha_n m_e}
\end{equation}
depends on the electron helicity $s=\pm 1$ and the quantity
$\varepsilon =\pm 1$ which splits the solutions into positive- and
negative-frequency branches.

\section{Spin light of electron in matter}

To the lowest order of the perturbation theory the corresponding
quantum process is described by the one-photon emission diagram
with the initial $\psi_{i}$ and final $\psi_{f}$ electron states
and with the vertex corresponding to the standard electromagnetic
interaction with the photon due to the electron charge $e$. Thus,
the transition amplitude is giving by
\begin{equation}\label{amplitude_e}
   S_{f i}=-ie \sqrt{4\pi}\int d^{4} x {\bar \psi}_{f}(x)
  ({\gamma}^{\mu}{e}^{*}_{\mu})\frac{e^{ikx}}{\sqrt{2\omega L^{3}}}
   \psi_{i}(x),
\end{equation}
where $k^{\mu}=(\omega, \mathbf{k})$ and ${e}^{\mu}$ are the photon momentum
and polarization vector, respectively. Choosing the three-dimensional
transversal gauge and performing integration over the time variable, we get
\begin{equation}\label{amplitude_integrated}
   S_{f i}=
  i e {\sqrt {\frac {2\pi}{\omega L^{3}}}}
  2\pi\delta(E-E'-\omega)
  \int d^{3} x {\bar \psi}_{f}({\bf r})( \bm {\gamma }{\bf e}^{*})
  e^{i{\bf k}{\bf r}}
   \psi_{i}({\bf r}),
\end{equation}
The $\delta$-function in the last expression stands for the energy conservation
law with $E$ and $E'$ being energies of the initial and final neutrino states.
Performing further integration over the spatial coordinates we get also the
momentum conservation law, which together with the energy conservation law,
\begin{equation}\label{E_law}
    E=E'+ \omega, \quad \mathbf{p}=\mathbf{p'}+\mathbf{k},
\end{equation}
leads to the only possible transition of the electron when its chirality
changes from $s_i=-1$ to $s_f=1$. The corresponding photon energy then is given
by the expression \cite{StuJPA06}:
\begin{equation}\label{omega_e}
   \omega=\frac{2\alpha_n m_e p[\tilde{E}- (p+\alpha_n m_e)\cos\theta]}
   {(\tilde{E}-p\cos\theta)^2-(\alpha_n m_e)^2},
\end{equation}
where $\theta$ is the angle between the directions of initial neutrino and
emitted photon propagation. We also use the following notation
$\tilde{E}={E}-c\alpha_n m_e$. In the case of relativistic electrons and small
values of the matter density parameter $\alpha_n$ the photon energy is
\begin{equation}\label{omega_2}
    \omega_{SLe}=
    \frac {1}{1-\beta_e \cos
    \theta }\omega_0,\ \ \
\omega_0= \frac {G_{F}} {\sqrt{2}}n_n\beta_e,
\end{equation}
here $\beta_e$ is the electron speed in vacuum. From this expressions we
conclude that for the relativistic electron the energy range of the $SLe$ may
even extend up to energies peculiar to the spectrum of gamma-rays. We also
predict the existence of the electron-spin polarization effect in this process.

\section{Rate and power of the radiation}

With the expressions for the amplitude
(\ref{amplitude_integrated}) and the emitted photon energy
(\ref{omega_e}) we arrive to general formulas for the total rate
and power of the radiation:
\begin{equation}\label{Gamma_with_Int}
%   \Gamma = \frac{e^2}{2}{\int_0}^{\pi} \big( 1-\tilde{\beta}_e' \tilde{\beta}_e
%    - \frac{{m_e}^2}{\tilde{E}' \tilde{E}}
%    \big)(1-y \cos\theta)\frac{\omega \sin\theta}{1+\tilde{\beta}_e'
%    \, y} d\theta,
   \Gamma = \frac{e^2}{2}{\int_0}^{\pi} \frac{\omega
   }{1+\tilde{\beta}_e' \, y} \left( 1-\tilde{\beta}_e'
\tilde{\beta}_e - \frac{{m_e}^2}{\tilde{E}' \tilde{E}} \right)
(1-y \cos\theta) \sin\theta \, d\theta,
\end{equation}
\begin{equation}\label{I_with_Int}
%   \mathrm{I} = \frac{e^2}{2}{\int_0}^{\pi} \big( 1-\tilde{\beta}_e' \tilde{\beta}_e
%    - \frac{{m_e}^2}{\tilde{E}' \tilde{E}}
%    \big)(1-y \cos\theta)\frac{\omega^2 \sin\theta}{1+\tilde{\beta}_e'
%    \, y} d\theta,
   {\mathrm I} = \frac{e^2}{2}{\int_0}^{\pi} \frac{\omega^2
   }{1+\tilde{\beta}_e' \, y} \left( 1-\tilde{\beta}_e'
\tilde{\beta}_e - \frac{{m_e}^2}{\tilde{E}' \tilde{E}} \right)
(1-y \cos\theta) \sin\theta \, d\theta,
\end{equation}
where
%\begin{equation}\label{S_e}
%    S_e = \left( 1-\tilde{\beta}_e'
%\tilde{\beta}_e - \frac{{m_e}^2}{\tilde{E}' \tilde{E}} \right)
%(1-y \cos\theta).
%\end{equation}
%ідесь бvли введенv следуіие величинv, описvваіие групповуі
%скорость волновой функции электрона:
\begin{equation}\label{beta_e}
   \tilde{\beta}_e=\frac{p+\alpha_n m_e}{\tilde{E}}, \
   \tilde{\beta}_e'=\frac{p'-\alpha_n m_e}{\tilde{E}'},
\end{equation}
are the quantities, describing the initial and final electron
group velocities, and
\begin{equation}
\begin{array}{c}
    {E'}=E-\omega, \quad p'=K_e \, \omega - p,
\\
    K_e={\displaystyle \frac{\tilde{E}- p\cos\theta}{\alpha_n m_e}}, \ \ \
      {\displaystyle y=\frac{\omega-p \cos\theta}{p'}}.
\end{array}
\end{equation}
Taking integration in (\ref{Gamma_with_Int}) and
(\ref{I_with_Int}) we obtain:
\begin{multline}\label{Gamma_tot}
    \Gamma=e^2\frac{m_e^2 \Big[(1+2\alpha_n^2)m_e^2+2p^2\Big]}{4 p^2 (4\alpha_n p+m_e)^2 \sqrt{(p+m_e
    \alpha_n)^2+m_e^2}}
\\
    \times\left[(4\alpha_n p+m_e)^2 \ln\big(1+\frac{4 \alpha_n p}{m_e}\big)
    -4\alpha_n p \, (m_e+6\alpha_n p)\right]
\end{multline}
and
\begin{multline}\label{I_tot}
    {\mathrm I}=e^2 \frac{m_e^2}{2p^2(4\alpha_n p+m_e)^3}
    \bigg\{
       \Big((1+\alpha_n^2)m_e^2+p^2\Big)(4\alpha_n p+m_e)^3 \ln\Big
       (1+\frac{4 \alpha_n p}{m_e}\Big)
       \\
       -\frac{4}{3}\alpha_n p
       \bigg[
          88\alpha_n^2 p^4+3 (1+\alpha_n^2) m_e^4+30
          \alpha_n (1+\alpha_n^2) m_e^3 p
          \\
          +p^2 m_e^2\big(3+88\alpha_n^2(1+\alpha_n^2)\big)+2\alpha_n(15+16\alpha_n^2)m_e p^3
       \bigg]
    \bigg\}.
\end{multline}

Let us estimate the total rate $\Gamma$ of the radiation, and also the
corresponding life-time $T_{SLe}$ of the electron in respect to the considered
process  using expression  (\ref{Gamma_tot}). Consider the electron with
momentum $p= 1 \ MeV$ moving in matter characterized by  the number density
$n_n \sim 10^{37} \ \text{cm}^{-3}$, in this case the matter density parameter
is $\alpha_{n}= 0.6\times 10^{-6}$. Then, for the rate of the process we get
$\Gamma \sim 3.2 \times 10^{-10} MeV$ which corresponds to the characteristic
life-time of the electron $T_{SLe}\sim 2\times 10^{-2} s$.

Finally, we have developed the approach to description of the matter influence
on an electron which is based on the exact solutions of the correspondent
modified Dirac equation for the particle wave function.  The approach developed
(we have previously used it for the case of neutrino) is similar to the Furry
representation in quantum electrodynamics. Note that our focus has been on the
standard model interactions of electrons with the background matter. A similar
approach, which implies the use of the exact solutions of the correspondent
modified Dirac equations, can be developed in the case when electrons interact
with different external fields predicted within various extensions of the
standard model (see, for instance, \cite{CollKosPRD98, ZhuLobMurPRD06} and the
paper of V.Zhukovsky et al in this book). We have predicted and investigated in
some detail a new type of electromagnetic radiation (the spin light of electron
in matter, $SLe$) that can be emitted by the electron due its magnetic moment
within the standard model of interaction with the background matter. The
obtained $SLe$ energy spectrum shows that for ultra-relativistic electrons it
can even extend up to energy range peculiar to the spectrum of gamma-rays.
Comparing the rates of the spin light of neutrino and spin light of electron in
matter, we have predicted (see also \cite{StuJPA06}) that the latter is more
effective then the former. We have predicted also that the $SLe$ emitted by
ultra-relativistic electrons moving in dense astrophysical and cosmological
media can have consequences accessible for experimental observations.

%\newpage
%\input{Graph.tex}
\end{document}